\documentclass{elsart}
\usepackage{amssymb}
\usepackage{amsmath}
\usepackage[dvips]{graphicx}

\journal{: \quad  April 9, 1973, in French
version.\qquad\qquad\qquad\qquad\qquad\qquad\qquad\qquad\qquad\qquad\qquad\qquad\qquad\qquad\qquad\qquad\qquad\qquad\qquad
}
\input{tcilatex}
\begin{document}

\begin{frontmatter}

\title{Compressible Fluids:\\ The  discontinuity  of the vorticity vector on a shock
 wave in thermodynamical variables
 \\
  \textbf{\footnotesize\emph{ Translation    of \
C. R. Acad. Sci. Paris t. 276, A, p. 1377-1380 (1973)}}}

\author{Henri Gouin}
\ead{henri.gouin@univ-cezanne.fr}

\address {
 Universit\'e d'Aix-Marseille \&   C.N.R.S.  U.M.R. 6181, \\ Case 322, Av. Escadrille
 Normandie-Niemen, 13397 Marseille Cedex 20 France}

\begin{abstract}
The discontinuity   of the vorticity is written as a function of the
vector $ T \func{grad} s$, (where $T$ is the temperature and $s$ the
specific entropy). The expression is obtained thanks to potential
equations and independently of the mass conservation and the
equation of momentum balance.
\end{abstract}


\end{frontmatter}

\section{Introduction}

The aim of this note is to prove that, in the most general
instationary case of perfect compressible fluids, across a shock
wave we have the relations:
\begin{equation}
\begin{array}{c}
\left[ u({\func{rot}\,\mathbf{v}})_{tg}\right] ={\mathbf{n}}\wedge
\left[ T\
{\func{grad}\ s}\right] \\
\left[ ({\func{rot}\,\mathbf{v}})_{n}\right] \mathrm{=0}%
\end{array}%
\end{equation}%
where ${\mathbf{v}}$ is the velocity vector of the fluid, $u$ is the
fluid velocity with respect to the shock wave, $T$ is the
temperature, $s$ is the specific entropy, the indices $tg$ and $n$
indicate the tangential and normal components to the shock wave of
the vector $\func{rot}\,\mathbf{v}$ and the discontinuity of a
tensorial quantity $\alpha $ is denoted by $\left[ \alpha \right] $
(see[1]).

We take into account the following shock conditions:
\begin{equation}
\left[ \mathbf{v}_{tg}\right] =0
\end{equation}%
\begin{equation}
\left[ \frac{1}{2}\,u^{2}+h\right] =0
\end{equation}%
where $h$ denotes the specific enthalpy ($\displaystyle dh=Tds+\frac{dp}{%
\rho }$).

We do not use the following shock conditions:
\begin{equation}
\left[ \rho u\right] =0
\end{equation}
\begin{equation}
\left[ p+\rho u^{2}\right] =0
\end{equation}
where $\rho $ is the density and $p$ the pression of the fluid.

We will use the \emph{potential equations }obtained by P.\ Casal [2]
or J.\ Serrin [3] and expressing another form of the equations of
the motions of compressible perfect fluids.

The motion of a compressible fluid is represented by a continuous
mapping of a reference three-dimensional space $D_{o}$ in the
physical space $D_{t}$ occupied by the fluid at time $t$:
\begin{equation*}
\mathbf{x}=\varphi _{t}(\mathbf{X}),\ \mathbf{x}\in D_{t},\
\mathbf{X}\in D_{o},
\end{equation*}%
or equivalently by a continuous mapping $\Phi :W_{o}\longrightarrow
W,\ \mathbf{z}=\Phi (\mathbf{Z})$ where $W_{o}$ is a
four-dimensional reference space and $W$ the physical time-space,
\begin{equation*}
\mathbf{z}=\left(
\begin{array}{c}
t \\
\mathbf{x}%
\end{array}%
\right) \in W\text{ \ and \ }\mathbf{Z}=\left(
\begin{array}{c}
t \\
\mathbf{X}%
\end{array}%
\right) \in W_{o}\,.
\end{equation*}%
We assume the motion has a shock wave localized on a surface $S(t)$
propagating in $D_t$, image by $\varphi _{t}$ of a  surface
$S_{o}(t)$ propagating in $D_{o}$. We denote by $\mathbf{n_{o}}$ and
$\mathbf{n}$ the unit normal vectors to $  S_{o}(t)$\ and $S(t)$
respectively, and $g_{o}$\ and $g$ their respective velocities; then
$u=\mathbf{n}^{T}\mathbf{v}-g$, where $^{T}$ denotes the
transposition.

Equivalently, $\Sigma _{o}$\ and $\Sigma $ are respectively the two
corresponding surfaces propagating in $W_{o}\ $and $W$; $\mathbf{N_o}$ and $%
\mathbf{N}$ are the associated normal vectors, $$ \mathbf{N}^{T}\mathbf{=}%
(-g\mathbf{,n}^{T}\mathbf{)} .$$
Consequently, $\Phi $ is a differential mapping on $W_{o}$, except on $%
\Sigma _{o}$; its Jacobian matrix is denoted by $\partial \mathbf{z}%
/\partial \mathbf{Z}$, F denotes the Jacobian matrix of $\varphi _{t}$:%
\begin{equation*}
d\mathbf{x}=\mathbf{v}dt+Fd\mathbf{X}.
\end{equation*}

\section{Exterior derivative}

The covector $\mathbf{C}=\mathbf{v}^{T}$ has an inverse image
$\mathbf{C_o} $ in $D_{o}$ such that
\begin{equation*}
\mathbf{C}_{\mathbf{o}}=\mathbf{C}F
\end{equation*}%
The exterior derivative of the form $\mathbf{C}$ is a 2-form which
is isomorph to the vector $\func{rot}\,\mathbf{v}$. It the image of
the 2-form
which is the exterior derivative of $\mathbf{C_o}$ isomorph to the vector $%
\func{rot_o}\left(\mathbf{C_o}^T\right)$:%
\begin{equation}
\func{rot}\,\mathbf{v}=\frac{F}{\det \ F}\
\func{rot_o}\left(\mathbf{C_o}^T\right)
\end{equation}%
$\func{rot_o}$ is the rotational on $D_{o}$ (see reference[4]).

The discontinuity of the vorticity vector comes from to parts: one
part comes from the discontinuity of its image
$\func{rot_o}\left(\mathbf{C_o}^T\right)$, and the other part comes
from the discontinuity of the Jacobian $F$.

\section{Discontinuity of the Jacobian $F$}

$\partial \mathbf{z}/\partial \mathbf{Z}$ is a linear mapping
transforming any tangent vector to $S_{o}(t)$ in a tangent vector to
$S(t)$. If we denote by $\mathbf{n_o}^{\prime
}=-\mathbf{n_o}/g_{o,}$ we obtain:

\begin{equation*}
\left[ F\right] =\left[ \mathbf{v}\right] \mathbf{n_o}^{\prime
^{T}},\ \
\mathbf{n_o}^{\prime ^{T}}=\mathbf{n}^{T}\frac{F_{1}}{u_{1}}=\mathbf{n}^{T}%
\frac{F_{2}}{u_{2}}
\end{equation*}

where indices $1, 2$ indicate quantities upstream and downstream the
shock.
Consequently,%
\begin{equation*}
\mathbf{n}^{T}\left[ \frac{F}{u}\right] =0
\end{equation*}

Taking into account Eq. (2) we obtain
\begin{equation}
\left[ \mathbf{v}\right] =\left[ u\right] \mathbf{n,}
\end{equation}%
\begin{equation}
\left[ F\right] =\left[ u\right] \mathbf{n\,n_o}^{\prime ^{T}},
\end{equation}%
\begin{equation}
\left[ \frac{F}{\det \ F}\right] =\frac{\left[ u\right] }{u_{2}\det \ F_{1}}%
\left( \mathbf{n\,n}^{T}-\mathbf{I}\right) F_{1}.
\end{equation}

where $\mathbf{I}$ is the identity matrix.

\section{Discontinuity of the 1-form $\mathbf{C_o}$}

Due to the fact that $\mathbf{C_o}=\mathbf{v}^{T}F$ and by using
Eq.\ (7)
and Eq.\ (8), we obtain:%
\begin{equation}
\left[ \mathbf{C_o}\right] =\left[ u^{2}+gu\right]
\mathbf{n_o}^{\prime ^{T}}.
\end{equation}

\section{Discontinuity of potentials}

The dot  denotes the material derivative, $\Omega $ is the body
force
potential.\ We consider the two quantities $\varphi (t,\mathbf{X})$ and $%
\psi (t,\mathbf{X})$ (denoted potentials) such that
\begin{equation}
\dot{\varphi}=\beta (t,\mathbf{X}),
\end{equation}%
\begin{equation}
\dot{\psi}=\gamma (t,\mathbf{X}).
\end{equation}

$\beta $ and $\gamma $ are two scalar fields defined in each point
of the
flow and such that%
\begin{equation*}
\beta (t,\mathbf{X})=\frac{1}{2}\mathbf{v}^{2}-h-\Omega ,\ \ \gamma (t,%
\mathbf{X})=T.
\end{equation*}
There exists a covector $\mathbf{B}$ function only of $\mathbf{X}$ such that%
\begin{equation}
\mathbf{C}_{o}=\frac{\partial \varphi }{\partial \mathbf{X}}+\psi \frac{%
\partial s}{\partial \mathbf{X}}+\mathbf{B}.
\end{equation}
We can verify that Eq. (13) together with Eq. (11) and Eq. (12) are
equivalent to \textit{potential equations} proposed by P. Casal in
[2] and J.\ Serrin in [3].\\ With the condition of adiabaticity
$$\dot{s}=0,$$ and the equation of balance of mass $$\frac{\partial
\rho}{\partial t} +\func{div}( \rho\, \mathbf{v}) = 0,$$ we obtain
the complete set of motion equations.

We can choose $\varphi $ and $\psi $ null on the shock wave and
continuous through the shock surface in the following manner:
\begin{equation}
\varphi =\int_{f(\mathbf{X})}^{t}\beta (\tau,\mathbf{X})d\tau ,\ \ \
\ \psi =\int_{f(\mathbf{X})}^{t}\gamma (\tau,\mathbf{X})d\tau ,
\end{equation}
where $t=f(\mathbf{X})$ is the equation of the shock surface
$S_{o}(t)$ which is assumed regular.

Potential $\varphi $ being continuous through the shock surface,
\begin{equation*}
\left[ \frac{\partial \varphi }{\partial \mathbf{X}}\right] =\left[ \dot{%
\varphi}\right] \ \mathbf{n_o}^{\prime ^{T}}=\left[ \frac{1}{2}\mathbf{v}%
^{2}-h\right] \ \mathbf{n_o}^{\prime ^{T}}.
\end{equation*}
By using Eq. (2), Eq. (3) and Eq. (10) we get:
\begin{equation}
\left[ \frac{\partial \varphi }{\partial \mathbf{X}}\right] =\left[ \mathbf{%
C_o}\right].
\end{equation}
Due to the fact that $\psi$ is null on the shock wave, Eq. (13)
expresses $\mathbf{B}$ is continuous through the shock:
\begin{equation}
\left[ \mathbf{B}\right] =0
\end{equation}

\section{Discontinuity of the image of the vorticity}

Let us consider%
\begin{equation*}
\mathbf{W_o}=\mathbf{C_o}^{T}-\left( \frac{\partial \varphi
}{\partial
\mathbf{X}}\right) ^{T},\text{ \ \ then  \ \ \ }\func{rot_o}\mathbf{W_o}=%
\func{rot_o}\mathbf{C_o}^{T}.
\end{equation*}
Eq. (13) yields $\dot{\mathbf{W_o}}=T\func{grad_o}s$, the value of $\func{%
grad_o}s$ being defined on $W_{o}$.

Due to Eq. (15), $\mathbf{W_o}$ is continuous through the shock and we get:%
\begin{equation*}
\left[ \frac{\partial \mathbf{W_o}}{\partial \mathbf{X}}\right]
=\left[
\dot{\mathbf{W_o}}\right] \ \mathbf{n_o}^{\prime ^{T}},\ \ \ \left[ \func{%
rot_o}\mathbf{W_o}\right] =\mathbf{n_o}^{\prime }\wedge \left[ \dot{%
\mathbf{W_o}}\right],
\end{equation*}
\begin{equation}
\left[ \func{rot_o}\mathbf{C_o}^{T}\right] =\mathbf{n}_{o}^{\prime }\wedge %
\left[ T\func{grad_o}s\right] .
\end{equation}
The discontinuity of the image of the vorticity is only tangential.
Using the previous results and application $\Phi$, we can verify
that this property of the vorticity remains true.

\section{Discontinuity of the vorticity}

The results obtained by Eqs (6), (9) and (17) allow to obtain Formulae (1).%
\newline
This expression general for non stationary perfect compressible
fluids is different from the result given by Hayes [5]. This is due
to the fact the result is obtained thanks to Eq. (3) of conservation
of energy. It neither uses Eq. (5) of the balance of the quantity of
motion nor Eq. (4) of the conservation of mass. To
obtain Formulae (1), the knowledge of the enthalpy field is only necessary.%
\newline
In the special case of a stationary, iso-energetic, irrotational
motion upstream of the shock, the relation can be expressed with the
help of the curvature tensor of the shock surface [6].


\begin{thebibliography}{9}
\bibitem{1} J. Hadamard, Le\c cons sur la propagation des ondes et les
\'equations de l'hydrodynamique, Chelsea Publ., New York (1949).

\bibitem{2} P. casal, Journal de M\'ecanique, 5, n$^o$ 2, 1986, p. 149-161.

\bibitem{3} J. Serrin, Mathematical principle of classical fluid mechanics,
Fluid dynamics 1, Encyclopedia of Physics, VIII/1, Springer, New
York (1959).

\bibitem{4} H. Cartan, Calcul diff\'erentiel, Hermann, Paris (1967).

\bibitem{5} W.D. Hayes, Journal of Fluid Mechanics, 2, 1957, p. 595.

\bibitem{6} A.L. Jaumotte and P. Carri\`ere, Chocs et ondes de chocs,
Masson, Paris (1971), p. 67-68.
\end{thebibliography}
\end{document}